# A Data-driven Approach for Rapid Detection of Aeroelastic Modes from Flutter Flight Test Based on Limited Sensor Measurements

Arpan Das[a*], Pier Marzocca[a], Giuliano Coppotelli[b], Oleg Levinski[c], Paul Taylor[d]

[a]Sir Lawrence Wackett Defence and Aerospace Centre, RMIT University, Melbourne, Victoria, 3082, Australia

[b]Department of Mechanical and Aerospace Engineering, University of Rome, La Sapienza, Rome, Italy

[c]Defence Science and Technology Group, Fisherman's Bend, Victoria, 3207, Australia

[d]Dynamics, Gulfstream Aerospace Corporation, Savannah, Georgia, 31402, United States

**Abstract**

Flutter flight test involves the evaluation of the airframe's aeroelastic stability by applying artificial excitation on the aircraft lifting surfaces. The subsequent responses are captured and analyzed to extract the frequencies and damping characteristics of the system. However, noise contamination, turbulence, non-optimal excitation of modes, and sensor malfunction in one or more sensors make it time-consuming and corrupt the extraction process. In order to expedite the process of identifying and analyzing aeroelastic modes, this study implements a time-delay embedded Dynamic Mode Decomposition technique. This approach is complemented by Robust Principal Component Analysis methodology, and a sparsity promoting criterion which enables the automatic and optimal selection of sparse modes. The anonymized flutter flight test data, provided by the fifth author of this research paper, is utilized in this implementation. The methodology assumes no knowledge of the input excitation, only deals with the responses captured by accelerometer channels, and rapidly identifies the aeroelastic modes. By incorporating a compressed sensing algorithm, the methodology gains the ability to identify aeroelastic modes, even when the number of available sensors is limited. This augmentation greatly enhances the methodology's robustness and effectiveness, making it an excellent choice for real-time implementation during flutter test campaigns.

*Keywords: Time-delay embedded Dynamic Mode Decomposition; Flutter Flight Test; Limited Sensor Measurements; Compressed Sensing; Data-driven Real-time Mode Identification.*

**Nomenclature**

Corresponding author
*Arpan Das, Research Assistant, RMIT university, Melbourne, Victoria 3082, Australia
Email: arpand1989@gmail.com

| Symbol | | Description |
|---|---|---|
| $A$ | = | Linear operator |
| $\tilde{A}$ | = | Projection of $A$ onto POD modes |
| $b$ | = | Vector coefficients of DMD modes |
| $b_k$ | = | Initial amplitude of DMD modes |
| $b_{opt}$ | = | Optimal mode amplitudes |
| $\beta$ | = | Aspect ratio of snapshots |
| $C$ | = | Measurement matrix |
| $D_b$ | = | Diagonal matrix containing amplitude of modes |
| $\gamma$ | = | Sparsity value |
| $J(b)$ | = | Cost function of $b$ |
| $L$ | = | Low-rank component |
| $\Lambda$ | = | Eigen values matrix |
| $\lambda_0$ | = | Hyperparameter for Augmented Lagrange Multiplier |
| $m$ | = | Number of snapshots taken |
| $n$ | = | Spatial points per snapshot |
| $\Omega$ | = | Diagonal matrix of eigenvalues |
| $\eta$ | = | Noise magnitude |
| $\Phi$ | = | DMD Modes |
| $\Phi_Y$ | = | Compressed DMD modes |
| $r$ | = | Rank of the reduced SVD |
| $S$ | = | Sparse component |
| $\Sigma$ | = | Matrix containing singular values |
| $\tau$ | = | Optimal threshold |
| $U$ | = | Left Singular Vector |
| $\mu_\beta$ | = | Median Marcenko-Pastur distribution |
| $V$ | = | Right Singular Vector |
| $V_\lambda$ | = | Vandermonde matrix |
| $W$ | = | Eigen vector |
| $X$ | = | Data matrix |
| $\tilde{X}$ | = | POD projected data matrix |
| $Y$ | = | Subsampled data matrix |
| **DMD** | = | Dynamic Mode Decomposition |
| **POD** | = | Proper Orthogonal Decomposition |
| **RPCA** | = | Robust Principal Component Analysis |
| **SVD** | = | Singular Value Decomposition |

# 1. Introduction

Flutter is a phenomenon of dynamic instability experienced by an aircraft during flight, resulting from the interplay of aerodynamic, inertial, and elastic forces. This interaction triggers an energy exchange, which is evident in the fluctuation of the damping rate across two or more structural modes. Flow-induced structural motion is a significant cause for concern as it can lead to fatigue failure, and the CFR (Code of Federal Regulations) part 25, 25.629 requires that no instability is present within the flight envelope [1].

Flutter flight test is an expensive and dangerous endeavor. This involves some form of artificial excitation applied to the lifting surface of the airframe and the measurement of subsequent responses due to that excitation, which is followed by an assessment of the airframe's aeroelastic stability. In a flutter flight test campaign, for a particular test point, at a given Mach number and altitude, the pilot stabilizes the aircraft, then with the help of Flight Test Interface (FTI), inputs a control surface pulse separately into the aileron, elevator, and rudder. All the data, including excitation force, response acceleration, and various important flight parameters such as speed, altitude, fuel weight, and aircraft configuration are recorded electronically on-board and sent to the ground station for real-time analysis. Flutter analysts at the ground station monitor and ensure all responses are damped. All the maneuvers are done at this state, then the pilot takes the aircraft to a safe speed while analysts at the ground station analyze the damping characteristics from a series of accelerometer channels. After the point is cleared by the flutter telemetry crew, the pilot takes the aircraft to the next point of the test and repeats the procedure discussed above.

Data obtained from on-board accelerometers are sent to the ground station in limited quantities based on the available bandwidth. This data is analyzed to extract the damping characteristics from each test point. Unfortunately, noise contamination, turbulence, non-optimal excitation of modes, and sensor malfunction in one or more sensors complicate the system identification process, and makes it time-consuming to identify the aeroelastic modes.

The primary aim of conducting flight flutter testing is to ascertain the frequencies and damping rates at each test point. Since the 1980s, the research community has made significant progress in developing new techniques to enhance the efficiency of flutter prediction [2–7]. These articles [8–11] of the most effective and well-known flutter prediction methodologies. Furthermore, [12–14] provides flutter analysis and prediction methodologies. Typically, flutter flight tests involve the application of an artificial excitation force, enabling the determination of system dynamics through input/output analysis. However, when the input signal consists of atmospheric turbulence, the detection methodologies must rely solely on output data [15,16]. Among the various methodologies available, a few deserve special mention, the autoregressive moving-average (ARMA) method [17,18], the nonlinear autoregressive moving-average exogenous (NARMAX) [19,20], Least-Square fitting of Complex Frequency domain functions (LSCF) method [21,22], and moving block approach (MBA) [23]. Among them, ARMA is a notably effective and efficient choice [24]. On the other hand, a considerably newer approach named Dynamic Mode Decomposition [25,26], which is an output-only method that decomposes that dataset into spatiotemporal modes has been utilized in flutter flight test analysis [27–30]. The standard DMD [31] shares several similarities with ARMA. However, a modified version of DMD called Higher-Order Dynamic Mode Decomposition (HODMD) [32] has been utilized [27–29] to analyze flutter flight testing data. The results have indicated significant improvements over ARMA. The abovementioned HODMD-based methods require a manual selection of a threshold criterion to extract the dominant DMD modes [32]. This criterion value for extracting the dominant modes differs when considering different dynamical systems, might differ for different flutter test points, and also while conducting tests on another aircraft platform.

This research entails showcasing a framework that time-delay embedded DMD alongside Robust Principle Component Analysis (RPCA) filtering technique (for outlier detection and removal), and a sparsity promoting criterion for selecting optimally sparse modes [33]. The main objective is to rapidly and automatically detect aeroelastic modes from flutter flight test data. The entire process can be completed in less than 45 seconds for each test point on a laptop-class computer. The anonymized flutter flight test data used for this study was provided by the fifth author of this paper. Moreover, the developed framework is augmented with a compressed sensing DMD [34] algorithm to detect the dominant aeroelastic modes from limited sensor measurements which makes it extremely fast (between 32 to 38 seconds for each test point) and effective in overcoming the problem when there are only a few sensors available for data acquisition, and in case of the occurrence of sensor malfunction during flight tests. The remainder of the paper is organized as follows: Section II describes the dataset and maneuver identification. Section III briefly describes the mathematical modeling and its implementation onto the developed frameworks adapted for rapid detection of aeroelastic modes with all and limited sensor data respectively. Section IV first discusses the effectiveness of the developed strategy by comparing the current results with previous literature with the same dataset utilizing HODMD and LSCF-based methods [27–29]. Thereafter, the results obtained from the three flight tests are discussed. Finally, the effectiveness and limitations of the compressed sensing augmented framework with limited sensor measurements are showcased. Section V provides a summary and offers concluding remarks.

## 2. Dataset and Maneuver Identification

The dataset used in this study was provided by the fifth author. It contains anonymized accelerometer channel data from three test points of a flight flutter test conducted on a Gulfstream aircraft. Data was obtained from a total of 92 accelerometer channels, from various locations on the aircraft. However, it has been determined that out of the total of 92 sensors, five of them were defective, and generated non-numeric readings. Therefore, those were not taken into account for the calculations and the remaining 87 are used. Each flight test was conducted at different test points having some Mach number and altitude values. The dataset was acquired by capturing the data from six consecutive maneuvers, which were artificially generated by implementing six impulsive excitations on the control surfaces. This involved two excitations on the aileron, two on the elevator, and two on the rudder. Due to the proprietary nature of the data, the aircraft type is not known, and the details about the test points such as the Mach number and altitude values at which these three tests are conducted are not known. Moreover, the other additional details, such as the sampling rate, location of the sensors, speed, fuel weight, aircraft configuration, and excitation force details are not known. The only available information is the accelerometer channel data (sampling rate unknown) acquired during the three flight tests conducted having three different test points, and the fact that the captured accelerometer response was generated by six symmetric and antisymmetric impulsive excitations on aileron, elevator, and rudder.

Careful observation reveals that the dataset is quite noisy. In some sensors, the noise contamination level is too high as shown in Fig. 1a. It is assumed for the sake of representation of the plots that $\Delta t = 1s$, but all the frequency content represented in this paper is scaled by $2f\Delta t$ as the sampling rate is unknown, therefore choosing any arbitrary $\Delta t$ would result in the same value for $2f\Delta t$. Fig. 1b shows the power spectrum of that noisy signal and it can be easily identified that the fundamental frequency is buried within the noise. To get an overview of the noise level on various sensors used for the three test points conducted, the signal-to-noise (SNR) ratio is plotted in Fig. 2. The different maneuvers can be easily identified by observing the signals as shown in Fig. 3, where the damping patterns are illustrated by the dotted red box. The detection of peaks in the signal allows for the easy identification of the maneuvers being performed. However, the sixth maneuver, from most of the sensors is undetectable. This might be because the data is either cut off due to bandwidth requirements or is deemed unnecessary. For all subsequent calculations, it is important to note that only the first five maneuvers will be taken into account for our analyses.

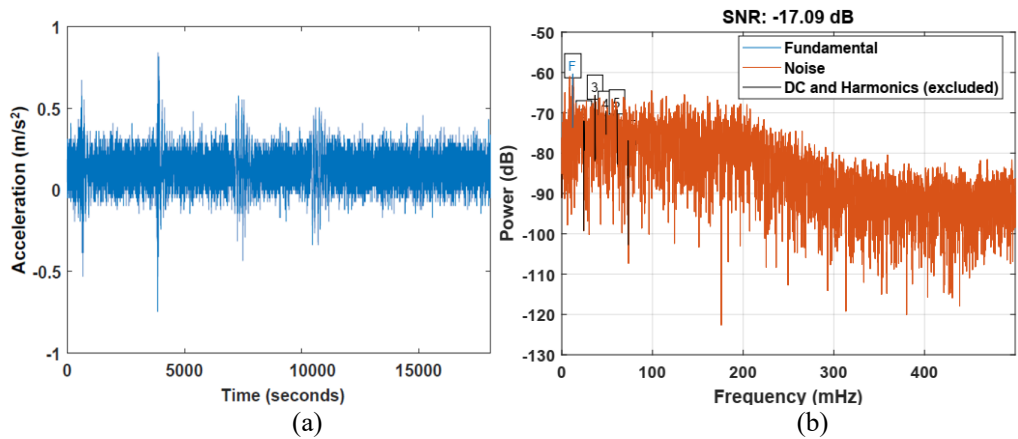

**Fig. 1**: (a) Noisy data from a sensor (b) power spectrum of the noisy signal

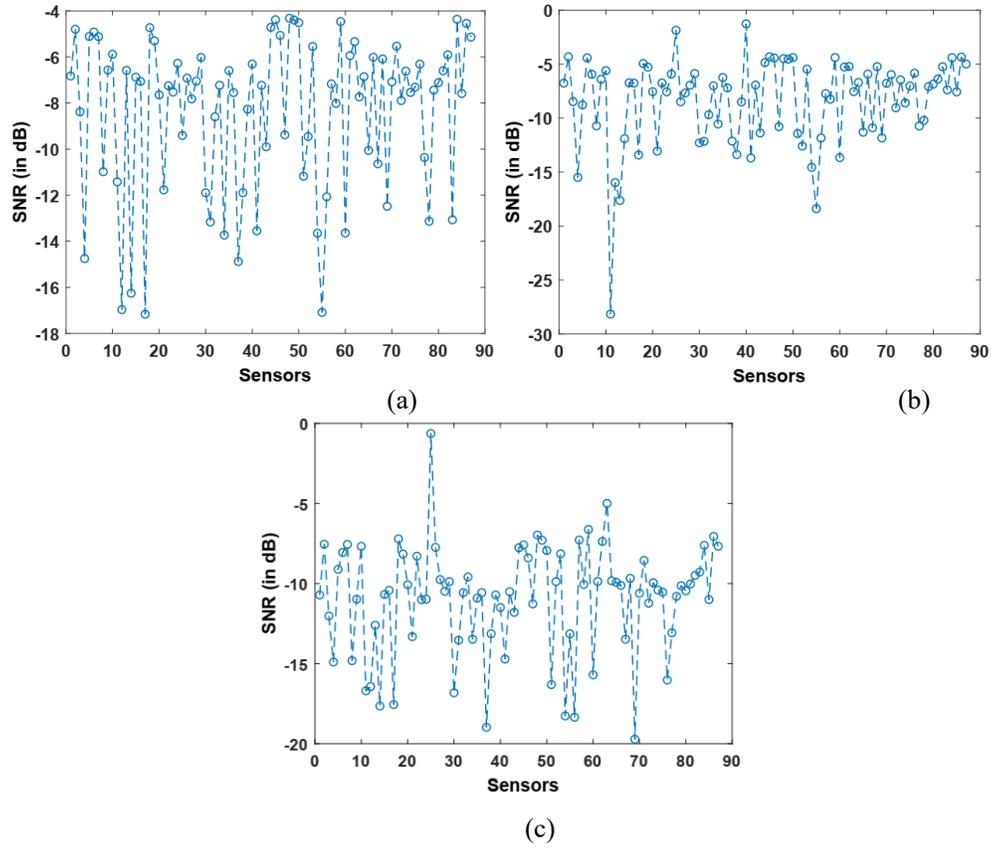

**Fig. 2**: Signal-to-noise ratio for (a) Test Point I (b) Test Point II (c) Test Point III

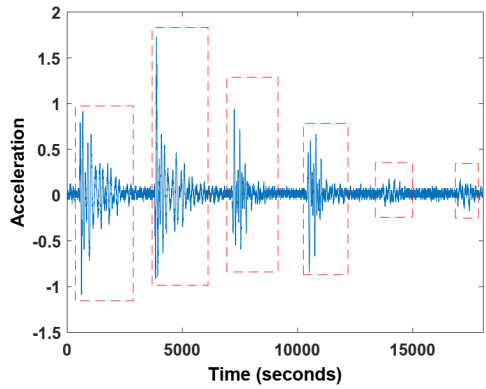

**Fig. 3**: Maneuver identification demonstrated with the signal captured from an accelerometer channel

## 3. Mathematical Modeling

The developed methodology to rapidly identify the aeroelastic modes requires a conjunction of methods including RPCA (Robust Principal Component Analysis) for outlier mitigation, Time-delay embedded DMD with Sparsity-promoting criterion for optimally sparse mode selection, and Compressed Sensing DMD to identify the modes based on limited sensor measurements. To be reasonably self-contained, these methods are discussed in brief.

*3.1 Time-Delay Embedded Dynamic Mode Decomposition with Sparsity-promoting Criterion*

The DMD method enables the decomposition of data into dynamic modes, which capture both spatial and temporal characteristics. These dynamic modes are derived from snapshots or measurements of a system recorded over a period of time. The extraction of dynamic modes from time-resolved snapshots is based on Arnoldi's algorithm [35–37]. The process of collecting snapshot data entails considering two variables: the quantity of snapshots captured (m) and the quantity of spatial points stored per snapshot (n). It was first discovered by Schmid [38,39] and is generally referred to as Standard DMD. A more modern definition called Exact DMD is given by Tu et al. [40]. The two algorithms are either practically the same or similar except for the DMD mode formulation, although they provide almost identical results. For the sake of conciseness, while addressing these methods, they are referred to as Standard DMD in this paper. In order to address the issue of the inability of standard DMD to capture standing wave, specific modifications to the methodology are necessary. In this case, time-delay embedding of the data matrix by creating a Hankel matrix is utilized to increase the rank of the data matrix [40]. Therefore, Takens' delay embedding theorem [41] is incorporated with the DMD algorithm. Furthermore, apart from capturing a standing wave, time-delay embedding plays a crucial role in addressing the nonlinear dynamics of a complex system. Indeed, the expansion resulting from time-delay embedding is suitable for representing different types of nonlinear dynamical systems, including attractors, dynamics deviating from attractors due to instabilities, and transient decay caused by attractors [32]. The procedure for performing Time-delay Embedded DMD with sparsity-promoting mode selection can be outlined as follows:

At first, a Singular Value Decomposition (SVD) of the entire data matrix [25,26] is computed. This is necessary because if the time-delay embedding of the data matrix is performed first, it would be computationally challenging to compute the SVD of $X$ in most cases, particularly when dealing with moderately high-dimensional measurements. Hence, the SVD is carried out on the entire set of measurements $X$, such that,

$$X \approx U\Sigma V^* \tag{1}$$

After computing SVD of the data, it is necessary to perform a rank truncation based on the computed singular values. The determination of this choice is influenced by multiple factors, including the source, quality of data, and the dynamic significance of low-energy modes. Among the various threshold criteria, Gavish and Donoho [42] have provided a theoretical framework for achieving optimal singular value truncation, even when there is additive white noise error, among the different threshold criteria. The threshold criterion is as follows,

$$\tau = \lambda(\beta)\sqrt{n}\eta \tag{2}$$

where $\beta = n/m$ is the aspect ratio of the data matrix and $\lambda(\beta)$ is,

$$\lambda(\beta) = \left(2(\beta + 1) + \frac{8\beta}{(\beta+1)+(\beta^2+14\beta+1)}\right)^{1/2} \tag{3}$$

Generally, the noise magnitude $\eta$ is unknown and estimated directly from the SVD of the data matrix. In that case, the optimal threshold $\tau$ is given by,

$$\tau = \omega(\beta)\sigma_{median} \tag{4}$$

where $\omega(\beta) = \lambda(\beta)/\mu_\beta$ and $\mu_\beta$ is the median of the Marcenko-Pastur distribution [42], respectively.

After the rank truncation, the data is projected onto its rank truncated Proper Orthogonal Decomposition (POD) modes, such that,

$$\widetilde{X} = U^*X = \Sigma V^* \tag{5}$$

The concept of Time-delay Embedded DMD and HODMD integrates the principles of Standard DMD with Takens' delay embedding theorem [41]. This integration allows for the capture of phase information related to a standing wave's eigenvalue pair. Additionally, time-delay embedding aids in enhancing the rank of the data matrix when the state measurement is of low dimensionality. This technique can be applied to both over-determined and under-determined data matrices. Consequently, the application of time-delay embedding to the reduced matrix can be represented as,

$$\widetilde{X}_{k+d} \approx \widetilde{A}_1\widetilde{X}_k + \widetilde{A}_2 X_{k+1} + \cdots + \widetilde{A}_d\widetilde{X}_{k+d-1} \tag{6}$$

The time-delayed reduced data-matrix $\widetilde{X}$ is divided into two sets $\widetilde{X}_1$ and $\widetilde{X}_2$, similar to standard/exact DMD algorithm, such as,

$$\widetilde{X}_2 = \widetilde{A}\widetilde{X}_1 \tag{7}$$

After that, another SVD of the reduced data-matrix $\widetilde{X}_1$ is calculated as,

$$\widetilde{X}_1 = U_1\Sigma_1 V_1^* \tag{8}$$

Consequently, another rank truncation is conducted based on some suitable criteria, and the reduced linear operator $\widetilde{A}$ computed as,

$$\widetilde{A} = \widetilde{X}_2\widetilde{X}_1^\dagger \tag{9}$$

$$\widetilde{A} = \widetilde{X}_2 V_1 \Sigma_1^{-1} U_1^* \tag{10}$$

The eigenvalue decomposition of $\widetilde{A}$ computes as,

$$\widetilde{A}W = W\Lambda \tag{11}$$

After the reconstruction of the eigen-decomposition of $A$ from $W$ and $\Lambda$, the eigenvalues are given by $\Lambda$ and the eigenvectors (reduced DMD modes) are represented by the columns of $\widetilde{\Phi}$:

$$\widetilde{\Phi} = \widetilde{X}_2 V_1 \Sigma_1^{-1} W \tag{12}$$

In the final step, the reduced DMD mode is projected back to full-order state (DMD mode) by,

$$\mathbf{\Phi} = U_1 \tilde{\mathbf{\Phi}} \qquad (13)$$

Thus, in order to derive the projected solution for all future time intervals, it is essential to utilize the low-rank approximation of eigenvalues and eigenvectors. For the sake of convenience, redefining $\omega_k = \ln(\lambda_k)/\Delta t$, it is possible to acquire an approximate solution for all future temporal states as,

$$x(t) \approx \sum_{k=1}^{r} \boldsymbol{\phi}_k \exp(\omega_k t)\, b_k = \mathbf{\Phi} \exp(\Omega t)\, \mathbf{b} \qquad (14)$$

where $b_k$ is the initial amplitude of each mode, $\mathbf{\Phi}$ is the matrix having columns as the DMD eigenvectors $\boldsymbol{\phi}_k$, and $\Omega = diag(\omega)$ is a diagonal matrix with eigenvalues $\omega_k$ as its diagonal entries. The eigenvectors $\boldsymbol{\phi}_k$ have the same size as the state vector $x$, and the coefficients $b_k$ are stored in the state vector $\mathbf{b}$. In order to calculate the value of $x(t)$, it is necessary to solve for the initial coefficient values $b_k$. The initial snapshot $x_1$ at time $t_1 = 0$ is defined as $x_1 = \mathbf{\Phi} \mathbf{b}$. It should be noted that the matrix of eigenvectors $\mathbf{\Phi}$ is not necessarily a square matrix. As a result, the vector $\mathbf{b}$ can be computed as follows:

$$\mathbf{b} = \mathbf{\Phi}^\dagger x_1 \qquad (15)$$

where $\mathbf{\Phi}^\dagger$ is the pseudoinverse of $\mathbf{\Phi}$.

The standard/exact DMD method involves the computation of the mode amplitude vector $\mathbf{b}$ by obtaining the pseudoinverse of $\mathbf{\Phi}$ and subsequently multiplying it with the first snapshot $x_1$. This approach is considered as the most direct way for computing $\mathbf{b}$. Nevertheless, it is essential to acknowledge that $x_1$ may not reside within the column space of $X'$, particularly in the presence of nonlinearity in the data. Hence, the identification of an optimal $\mathbf{b}$ assumes great significance, which is achieved through the minimization of a cost function $J(\mathbf{b})$,

$$\underset{b}{\text{minimize}}\, J(\mathbf{b}) = \|X - \mathbf{\Phi} D_b V_\lambda\|_F^2 \qquad (16)$$

where $D_b$ is a diagonal matrix that encompasses the amplitudes of the modes in its diagonal entries, and $V_\lambda$ represents a Vandermonde matrix that possesses eigenvalues. The product $\mathbf{\Phi} D_b V_\lambda$ is an alternative representation for $x(t)$ as in Eq. (14). By leveraging matrix trace properties, the function $J(\mathbf{b})$ can be expressed in a different form, such as,

$$J(\mathbf{b}) = \mathbf{b}^* P \mathbf{b} - q^* \mathbf{b} - \mathbf{b}^* q + s \qquad (17)$$

where, $P = (\mathbf{\Phi}^*\mathbf{\Phi}) \circ (\overline{V_\lambda V_\lambda^*})$, $q = \overline{diag\{V_\lambda X^* \mathbf{\Phi}\}}$, and $s = trace(\Sigma^*\Sigma)$. The objective function $J(\mathbf{b})$, in our case, is somewhat different from [43] as the definition of exact DMD is utilized to compute the reduced order DMD modes (Eq. 12), which in turn was used in deriving the cost function, whereas in the original paper [39,44] projected DMD $\mathbf{\Phi} = UW$ was used. The objective function is minimized at the point where its gradient vanishes. Consequently, the solution can be acquired analytically as,

$$\mathbf{b}_{optimal} = P^{-1}q \qquad (18)$$

The cost function is further enhanced by incorporating a sparsity structure, which facilitates the identification of a subset of DMD modes. This incorporation of sparsity enables the user to achieve the desired trade-off between the number of extracted modes and the approximation error. Consequently, the optimization problem is transformed into a more customizable and user-centric framework, such that,

$$\underset{b}{\text{minimize}}\, J(\mathbf{b}) + \gamma\, card(b) \qquad (19)$$

The $card(b)$ term penalizes non-zero elements in the vector of unknown magnitudes $\boldsymbol{b}$. The optimization problem is solved by incorporating $\ell_1$ penalty which promotes sparsity as shown in Eq. (20), and the user-defined value γ serves as an indicator of the extent of sparsity within the structure,

$$\underset{b}{\text{minimize}}\, J(\mathbf{b}) + \gamma \sum_{i=1}^{r} |b_i| \qquad (20)$$

The modified optimization problem presented in Eq. (20) is a convex optimization problem. To solve this problem, the Alternating Direction Method of Multipliers (ADMM) algorithm [45] is employed, resulting in the selection of the optimal set of mode amplitudes. To acquire a comprehensive understanding of the formulation and derivation of sparsity-promoting DMD, readers are encouraged to consult the work of Jovanovic et al. [43].

*3.2 Robust Principal Component Analysis*

Least-square regression is known to be extremely vulnerable to noise, outliers, and corrupted data. This vulnerability extends to other techniques like SVD and PCA, which are built upon the foundation of least square regression. Consequently, DMD, which relies on these techniques, is similarly affected by the presence of such problematic data. In order to mitigate this issue, Candes *et al.* [46] have introduced a technique that assumes the data matrix can be represented as a combination of two components: a low-rank and a sparse component, such that,

$$\boldsymbol{X} = \boldsymbol{L} + \boldsymbol{S} \qquad (21)$$

The process of decomposition involves splitting the data matrix X into two components: a low-rank component $\boldsymbol{L}$ and a sparse component $\boldsymbol{S}$. The sparse component S can contain noise, outliers, or missing data. By isolating the outliers and erroneous data in $\boldsymbol{S}$, the principal components of $\boldsymbol{L}$ remain robust. Hence, the ultimate objective is to find $\boldsymbol{L}$ and $\boldsymbol{S}$ in a manner that satisfies the following criteria,

$$\underset{L,S}{\min}\, rank(L) + \|S\|_0 \text{ subject to } L + S = X \qquad (22)$$

The evaluation of the sparsity of the data matrix involves quantifying the number of non-zero elements present within it, represented by $\|S\|_0$. It is important to mention that both the $rank(L)$ and the $\|S\|_0$ terms are not convex, making the problem computationally intractable. However, it is possible to achieve optimal solutions for $L$ and $S$ with a high probability by leveraging a convex relationship of Eq. (21), such that,

$$\min_{L,S} \|L\|_* + \lambda_0 \|S\|_1 \text{ subject to } L + S = X \tag{23}$$

The nuclear norm of the matrix $L$, denoted by $\|L\|_*$, is calculated by summing the singular values of the matrix $L$. On the other hand, the 1-norm of matrix $S$, denoted as $\|S\|_1$, is obtained by summing the magnitudes of each entry of $S$. This measure acts as a proxy for the zero norm of $S$. The hyperparameter $\lambda_0$ is obtained by $\lambda$ divided by the maximum value between $n$ and $m$, and is given by $\lambda_0 = \lambda/\sqrt{\max(n,m)}$, where $n$ and $m$ are the dimensions of the data matrix. It is important to note that the solution to Eq. (23) is likely to converge to the solution of Eq. (21) with a high probability when $\lambda$ is set to 1, under the assumption that $L$ is not sparse and $S$ is not of low rank. However, the validity of this condition may vary depending on the nature of the data matrix, which means that the value of $\lambda$ might need to be adjusted accordingly.

The Principal Component Pursuit (PCP) problem, represented by Eq. (23) is a convex optimization problem. It can be effectively addressed by employing the Augmented Lagrange Multiplier (ALM) algorithm, which was developed by Lin *et al.* [47]. The ALM technique operates using the augmented Lagrangian approach, which can be mathematically formulated as stated below,

$$\mathcal{L}(L,S,Y) = \|L\|_* + \lambda_0 \|S\|_1 + \langle Y, X - L - S \rangle + \frac{\mu}{2} \|X - L - S\|_F^2 \tag{24}$$

In order to address the PCP problem, one possible approach is to utilize a generic solution for the Lagrange multiplier algorithm. This involves finding the values of $L_k$ and $S_k$ that minimize the objective function $\mathcal{L}$, and subsequently updating the matrix of Lagrange multipliers by $Y_{k+1} = Y_k + \mu(X - L_k - S_k)$. This iterative process continues until the solution converges, as determined by a predefined threshold criteria. In the context of this particular study, the inexact Augmented Lagrange Multiplier algorithm, which was developed by by Lin *et al.* [47] is employed to solve the PCP problem.

*3.3 Compressed Sensing DMD*

If full-state measurements are available, it is possible to compress the data and compute DMD on the compressed data. Thereafter, the full-state DMD modes can be reconstructed by linearly combining full-state snapshots based on the compressed DMD transformations. This is computationally much more efficient than computing DMD on the full-state data directly. However, there are situations where full-state measurements are not available or acquiring full-state measurements is very expensive, in that case it is possible to reconstruct full-order DMD modes using compressed sensing.

It is possible to collect data $X, X'$ and compress data $Y, Y'$ such that, $Y = CX$, and $Y' = CX'$. Therefore, the best fit linear operator can be expressed as,

$$Y' = A_Y Y \tag{25}$$

where $C$ is the measurement matrix $\in \mathbb{R}^{p \times n}$, where $p$ is the number of measurements from the full-state data to create a subsampled data matrix. $A_Y$ is related to $A_X$, the linear operator from matrix from full-state measurements $X, X'$, such that,

$$CA_X = A_Y C \tag{26}$$

The following assumptions are made that the columns of $X, X'$ are sparse in a transform basis $\Psi$, so that $X = \Psi S$ and $X' = \Psi S'$, where $S$ and $S'$ have sparse columns. Also, measurements $C$ are incoherent with $\Psi$. Another assumption is that $X, X'$ is in the same sparse subset of the basis $\Psi$. This guarantees that the POD modes $U_X$ and the DMD modes $\phi_X$ are in the same sparse subspace as well. An important theorem that should be noted states that DMD eigenvalues are invariant to left transformations $C$ of data $X$ if $C$ is unitary, and the resulting DMD modes are projected through $C$, such as,

$$\phi_Y = C\phi_X = C\Psi \Phi_S \tag{27}$$

For more details on the assumptions, lemma, and proof of the theorems involved in this DMD algorithm, the readers are encouraged towards the paper by Brunton *et al.* [48].

In this work, the effectiveness of the developed framework, its efficacy, and limitations in rapid detection of the aeroelastic modes from subsampled or limited measurements are showcased. Therefore, we assume no knowledge of the full-state measurements, and we only have subsampled or compressed measurements $Y, Y'$. The details of the Compressed Sensing DMD algorithm are as follows,

$$Y \approx U_Y \Sigma_Y V_Y^* \tag{28}$$

Then the compressed linear operator is computed as,

$$\widetilde{A}_Y = U_Y^* Y' V_Y \Sigma_Y^{-1} \tag{29}$$

Subsequently, the eigen-decomposition of $\widetilde{A}_Y$ is computed, such that,

$$\widetilde{A}_Y W_Y = W_Y \Lambda_Y \tag{30}$$

where the columns of $W_Y$ are the eigenvectors and the diagonals of $\Lambda_Y$ contain the corresponding eigenvalues of the compressed or subsampled linear operator. Then the reduced order DMD modes are computed as,

$$\Phi_Y = Y' V_Y \Sigma_Y^{-1} W_Y \tag{31}$$

Finally, the full-state DMD modes can be computed from the reduced order DMD modes by performing $\ell_1$ minimization on $\phi_Y$ to solve $\phi_Y = C\Psi \phi_S$ for $\phi_S$, and then eventually $\phi_X$, since $\Phi_X = \Psi \Phi_S$. This is achieved either by using the Compressive Sampling Matching Pursuit (CoSaMP) algorithm [49], or by using Orthogonal Matching Pursuit (OMP) [50]. This is particularly useful if the spatial locations for the sensors (limited) are known, and can thereby be utilized for generating the aeroelastic mode shapes. In this case, due to the proprietary nature of the dataset, the spatial locations of the accelerometer channels are not known, therefore the present study is restricted to dominant aeroelastic mode identification and it's damping characteristics. For more details on these algorithms, the readers are directed towards [49,50].

## 4. Framework for Rapid Detection of Aeroelastic Modes

The initial and primary step in identifying aeroelastic modes through Dynamic Mode Decomposition (DMD) involves the creation of a data matrix that encompasses all the temporal snapshots. In this particular scenario, the flight test data is recorded in a temporally equispaced manner, although the exact time intervals are unknown due to the proprietary nature of the data. All the frequency content mentioned in this paper is the scaled version $2f\Delta t$. This is done for validation purposes in line with the previous literature that used the same dataset [27–29]. Moreover, choosing any arbitrary $\Delta t$ would result in the same values for $2f\Delta t$. As mentioned earlier, out of the six maneuvers, five will be taken into consideration for calculations, while the sixth one remains incomplete. The detection of these maneuvers can be achieved by observing the peaks in the signal. As a result, the signal from each sensor is segregated into separate maneuvers, which are subsequently consolidated in a matrix. Despite the length of the maneuvers, considering 2200 time-steps for each maneuver is deemed sufficient for our analyses. The entire process of stacking maneuvers from different sensors is visually depicted in Fig. 4. The figure illustrates how the five maneuvers from $Sensor_x$ are divided into segments and vertically stacked, followed by the stacking of maneuvers from $Sensor_y$, and so forth. The bottom five signals in the figure represent each segment from $Sensor_x$, corresponding to maneuvers 1 to 5. Therefore, for 87 sensors, there are 435 maneuvers, and each maneuver is 2200 time-steps long, which makes the data matrix of size 435x2200 which is rank deficient.

To address the issue of noise and outlier contamination, the data matrix undergoes a filtering process using the RPCA technique developed by Candes *et al.* [46]. Figure 5 showcases the efficacy of the RPCA filtration technique. Figures 5a-5d depict maneuvers captured by different sensors, with the plots showcasing both the original data with noise obtained directly from accelerometer channels and their corresponding RPCA-filtered counterparts. While the RPCA technique does not completely eliminate all noise contamination, it proves to be highly effective in reducing a significant portion of it. Das *et al.* [51] showcased RPCA technique in conjunction with DMD for salt and pepper type noise removal from corrupted buffet dataset to perform accurate modal diagnostics. Moreover, it demonstrates exceptional performance in rejecting outliers, as evidenced in Fig. 5d.

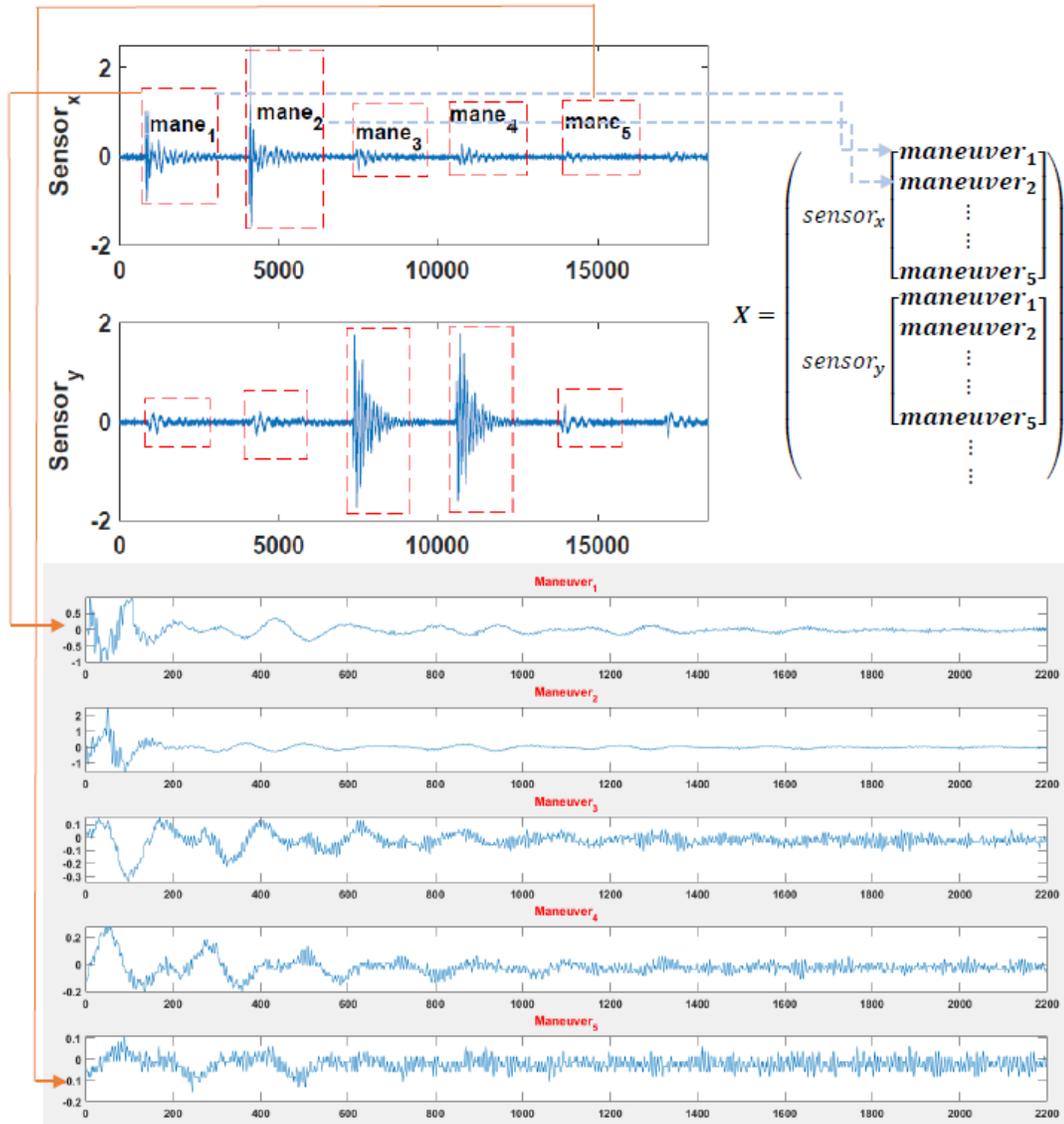

**Fig. 4**: Creation of data matrix involving maneuver stacking from different sensors

After the application of RPCA filtration, the subsequent step involves the computation of time-delay embedded DMD. In the initial stage of DMD computation, an SVD of the low-rank component of the data matrix is conducted to truncate the ranks at the first level. Following this, the data is projected onto the retained POD modes after rank truncation. Subsequently, time-delay embedding is performed on the projected data to generate a Hankel matrix based on the desired time-delay order. The eigenvalues and eigenvectors of the linear operator corresponding to the projected data matrix, utilizing the DMD methodology, are then computed (at this stage, another rank truncation can be executed). The full-state projection is achieved by multiplying the eigenvectors with the POD modes computed at the initial level of rank truncation. The reconstruction process involves the computation of DMD modes, their amplitudes, and frequencies. Furthermore, a reconstruction error based on the Relative RMS error is calculated. Once the reconstruction error reaches a certain threshold value, the computation is stopped. Otherwise, the entire process is repeated on the reconstructed data matrix until the error value is below the threshold limit. After meeting the convergence criteria for the reconstruction error, the sparsity promoting mode selection criterion is implemented to accurately identify the dominant aeroelastic modes from a large number of modes. The entire process can be comprehended by referring to the flowchart depicted in Fig. 6a.

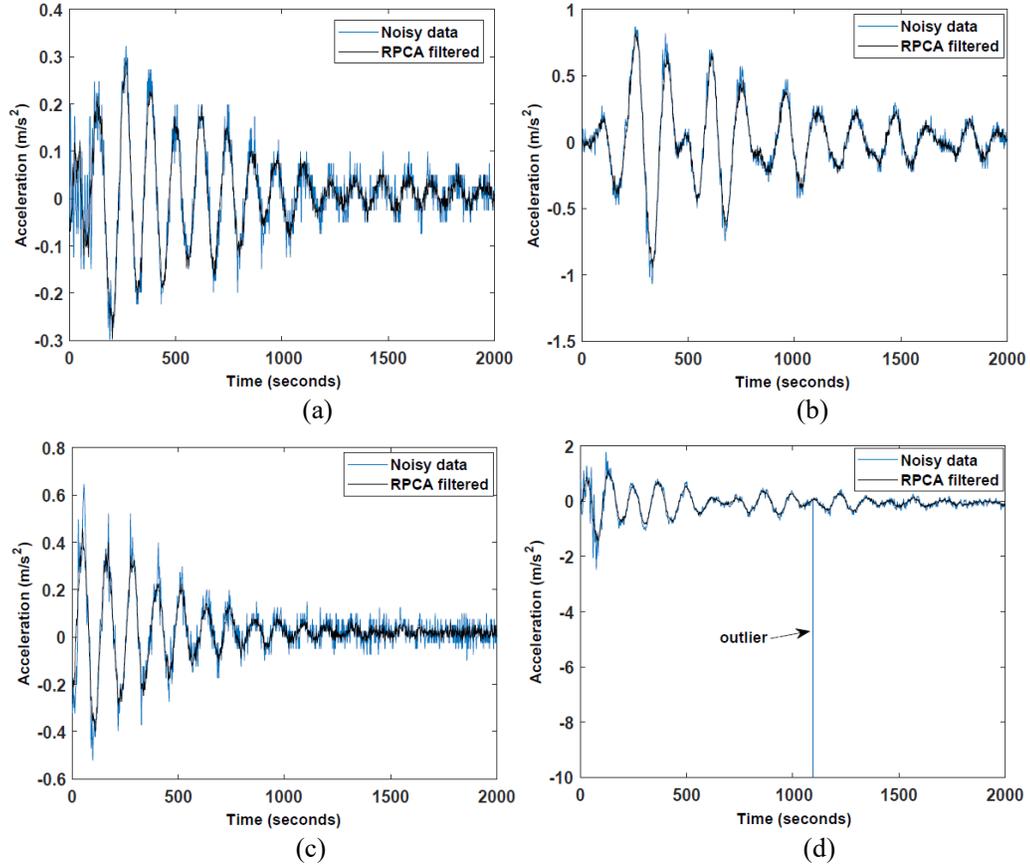

**Fig. 5**: Comparison of RPCA filtered data with its noisy counterpart from accelerometer channels

There are circumstances where sensor measurements are limited. These might be due to limited sensor placement because of cost reduction, physical constraints in putting more sensors, or might be due to the occurrence of sensor malfunction during testing. In that case, the framework developed in this work is illustrated with the help a flowchart. Fig. 6b illustrates the further modification of the framework employed to rapidly identify the aeroelastic modes when only a limited set of sensor measurements are available. While a significant portion of the framework is similar, some key changes distinguish the two frameworks. At first, we start with limited sensor measurements, meaning the knowledge of the full-state system with more sensors is not known in-priori. After that, up to the steps where time-delay embedding is applied for Hankel matrix creation, DMD computations on the Hankel matrix are the same. However, the key distinction starts after the DMD computations, as we started with limited measurements, the DMD modes computed are the compressed-sensing DMD modes, not the full-state DMD modes based on all the sensors. When the spatial location of the sensors is known, the mode shapes can be calculated using compressed sensing-based algorithms, such as Orthogonal

Matching Pursuit (OMP) or Compressive Sampling Matching Pursuit (CoSaMP) [49,50]. However, in the present study, the spatial location is unknown due to the proprietary nature of the data. In this scenario, the DMD reconstruction and reconstruction error are computed in relation to the limited measurement data matrix. When the reconstruction error satisfies the convergence criteria, the sparsity-promoting criterion [33] is utilized to identify the dominant aeroelastic modes by extracting the sparsely optimal modes. Even if the sensor locations are known, the subsequent step for mode selection remains unchanged and consistent. The major differences from the first framework are highlighted by the dotted green rectangular window illustrated in Fig. 6b.

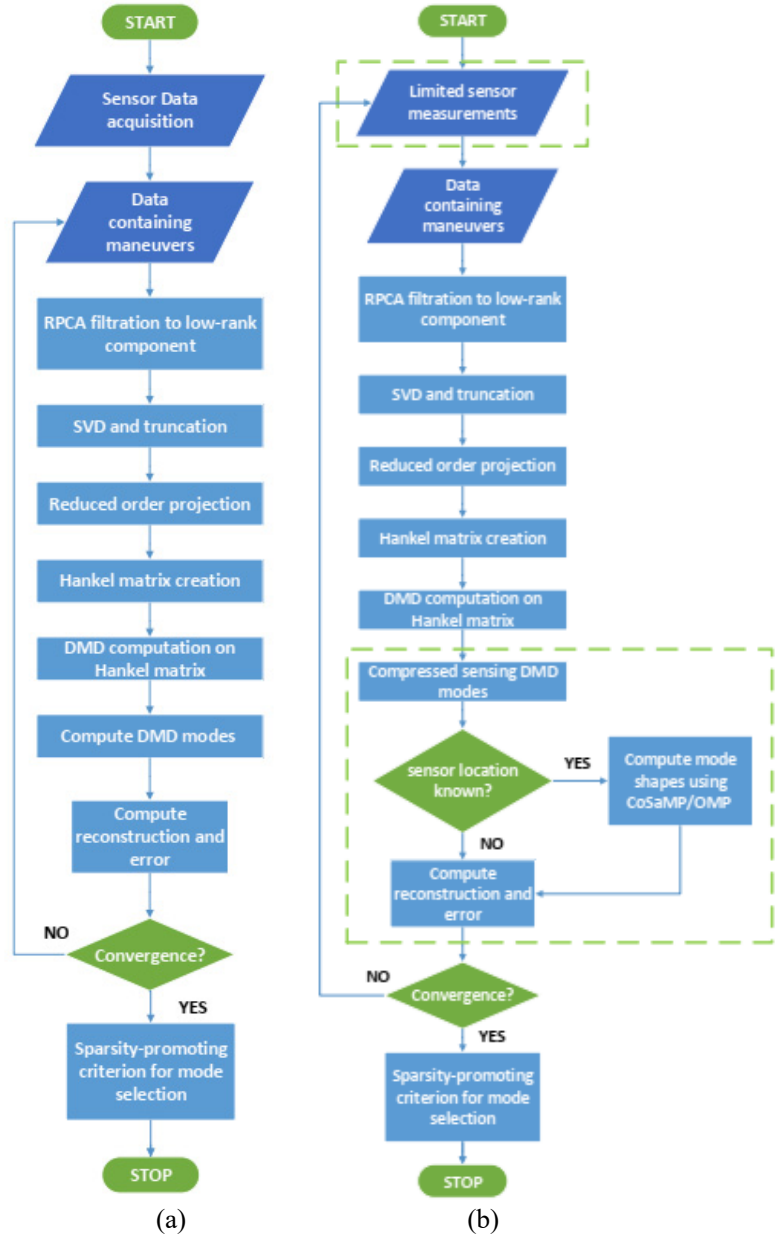

**Fig. 6**: Framework for rapid detection of aeroelastic modes based on (a) all sensors (b) limited sensors

# 5. Results and Discussion

In this study, the proposed methodology is applied to anonymized data from three flight test points. Each test point had six maneuvers (although the sixth was neglected for each test point) and a total of 92 sensors (of which 5 were ignored due to data corruption). Each maneuver was either an aileron, elevator, or rudder pulse. This section is further divided into three sub-sections, the first one discusses the thorough implementation of the developed algorithm on test point I and validates its efficacy by comparing the results from previous literatures with the same dataset [27–29]. The second one discusses results obtained from test points II and III. The third and final sub-section discusses the developed approach and its limitations for handling data with limited measurements to identify the aeroelastic modes with high accuracy.

*5.1 Flight Test Point I Results and Validation*

After the collection of data and its division into distinct maneuvers, the maneuvers are then combined to form a data matrix. This matrix is then subjected to filtering using the RPCA algorithm. Subsequently, the dataset is truncated to achieve order reduction. To do that, an SVD of the data matrix is taken and rank truncated to 101 modes considering the Gavish and Donoho criterion [52] as well as the fact that 99.9% of energy is kept by the first 101 modes (Fig. 7a). Nevertheless, it is crucial to recognize that even after implementing the RPCA algorithm to eliminate noise, the dataset remains significantly contaminated. As a result, a notable proportion of erroneous modes will inevitably be present within the set of 101 modes. Therefore, it becomes imperative to implement further truncation and mode selection criteria following the DMD computation to accurately identify the dominant aeroelastic modes.

The subsequent step involves projecting the data onto its truncated POD modes before proceeding to the creation of the Hankel matrix, as outlined in the flowchart depicted in Fig. 6a. The selection of the time-delay parameter (d) is determined by evaluating the reconstruction error of the Dynamic Mode Decomposition (DMD) computed data matrix in comparison to the original noisy data. This evaluation is performed by plotting the error with respect to the number of shift-stacking of rows or the time-delay parameter (d), as illustrated in Fig. 7b. From Fig. 7b, it is evident that a value of d=300 is sufficient for our analysis. This is because the Relative Root Mean Square error plot levels off beyond 300, and the difference in error between d=300 and d=400 is negligible (less than 1%). It is worth noting that the reconstruction error in this case exhibits large values and barely drops below 15%. This is due to the fact that the DMD reconstructed data matrix is compared with the noisy data, and the DMD reconstruction is performed after filtering the data matrix using the Robust Principal Component Analysis (RPCA) technique.

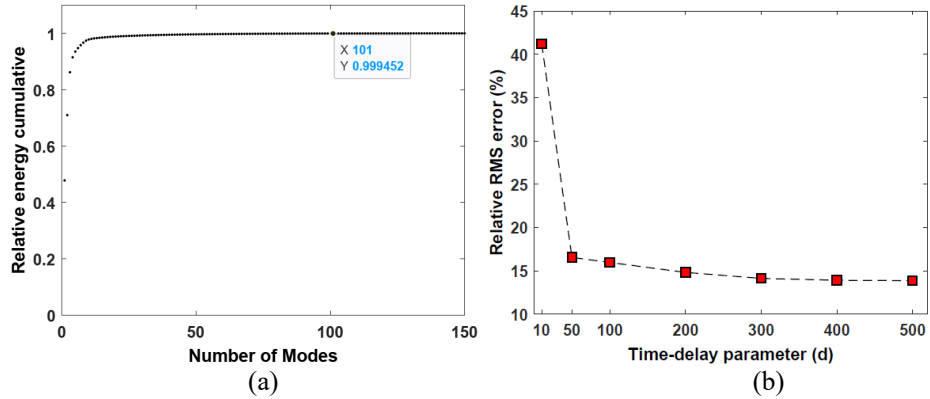

**Fig. 7**: (a) SVD rank truncation (b) Comparison of Relative RMS error w.r.t. time-delay parameter (d) (test point I)

According to the flowchart depicted in Figure 6a, the time-delayed Dynamic Mode Decomposition (DMD) is computed on the data matrix after setting the time-delay parameter. This computation is repeated until the Relative RMS error falls below a specific threshold. Once this condition is met, the sparsity-promoting DMD-based mode selection criterion [33] is employed to identify the dominant aeroelastic modes from the large number (101) of modes that were retained after truncating the data matrix following the SVD computation. To apply the sparsity-promoting criterion, a set of 200 sparsity ($\gamma$) values are chosen, which are logarithmically spaced between two extreme values. The minimum value identifies all the modes, meaning that all 101 mode amplitudes have non-zero values, while the other extreme value yields a two-mode solution. To determine the optimal sparsity value that results in the best sparse subset of modes, two factors must be taken into account: the variation in the number of non-zero modal amplitudes with respect to $\gamma$ (Fig. 8a), and the variation in performance loss (DMD reconstruction error) with respect to $\gamma$. (Fig. 8b). As expected, increasing $\gamma$ values lead to a smaller number of modes, but at the same time, the performance loss increases as the model becomes sparser. Therefore, to select the subset of sparsely optimal mode amplitudes and to identify the dominant aeroelastic modes, it is crucial to maintain the parsimony of the model. This involves striking a balance between the number of modes and the reconstruction error, so that an sparsely optimal solution, which accurately represents the dominant aeroelastic modes can be obtained.

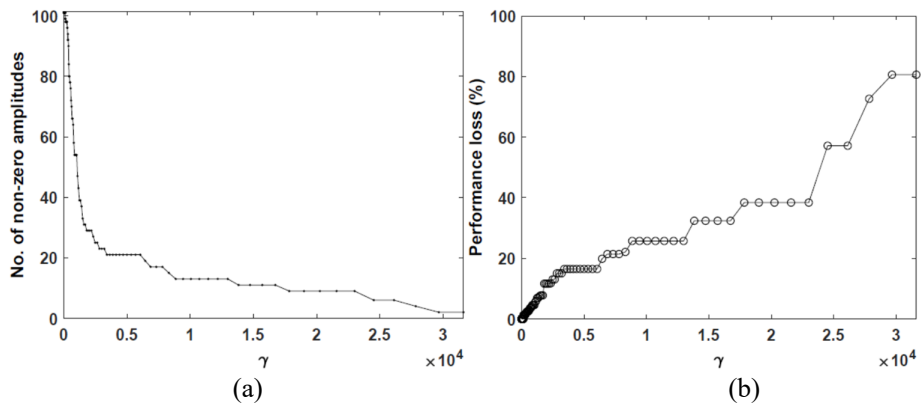

**Fig. 8**: Comparison between (a) Number of non-zero amplitudes Vs $\gamma$ (b) Performance loss Vs $\gamma$

Hence, it is crucial to conduct a comparison of both plots (Fig. 9) in order to determine the intersection point where the patterns align, indicating the most optimal level of sparsity. This level of sparsity corresponds to the set of modal amplitudes that are optimally sparse, with a $\gamma$ value of 6045.1 in this specific case. The process of identifying these sparsely optimal modes is automated and does not require any intervention from the user. Based on the determined $\gamma$ value, only 21 modes exhibit non-zero amplitudes. Since DMD modes are complex conjugates of each other, there are 10 distinct mode pairs and one individual mode, which is likely a static mode resulting from a potential offset in the time-history data. Subsequently, it becomes necessary to examine the DMD spectrum of these modes selected using a sparsity-promoting criterion, and comparing them to all the 101 modes if this criterion was not implemented, as depicted in Fig. 10. The DMD spectrum x-axis represents the real part of DMD eigenvalues related to the frequencies of the modes, and the y-axis represents the imaginary part of the eigenvalues pertaining to the growth rate of the modes, meaning modes inside the unit circle represent stable behavior, and modes outside the unit circle have positive growth rate meaning unstable behavior.

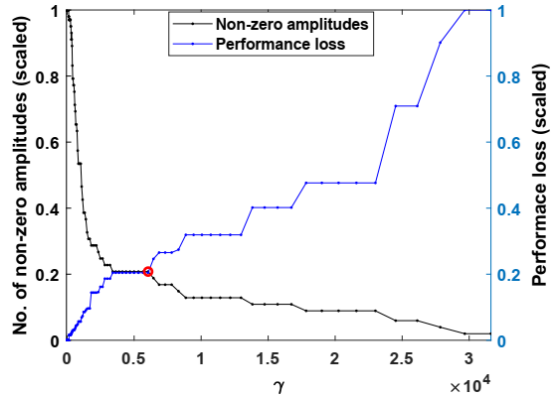

**Fig. 9**: Selection of optimal sparsity value based on sparsity structure and performance loss (Test point I)

The advantage of employing a sparsity-promoting criterion becomes apparent through the mode selection procedure. The group of modes extracted using the sparsity-promoting criterion has excellent stability and very low frequency which is ideal for extracting the dominant aeroelastic modes as shown in Fig. 10a. Moreover, the decay rate compared to frequency (scaled by $2f\Delta t$) of the selected modes shown by blue circles in Fig, 10a is plotted in Fig. 10b. As discussed previously, due to the proprietary nature of the dataset, the sampling rate is not known, that means we don't have any knowledge about the time-step size. Therefore, unit time-steps have been assumed and the obtained frequency is scaled by $2f\Delta t$, in line with the previous literature that used the same dataset [27–29] to compare with their results for validation purposes. Upon careful examination of Figure 10b, it becomes evident that the majority of the modes exhibit minimal damping, with the exception of a single mode positioned precisely at the zero line (static mode), which is probably due to DC offset and can be ignored. During flutter flight testing, due to the analytical margin requirement of 15%, the flight test campaign should not come close to flutter and some modes are lightly damped, which is expected. Moreover, some modes are not easily excited, especially with control surface pulses [53].

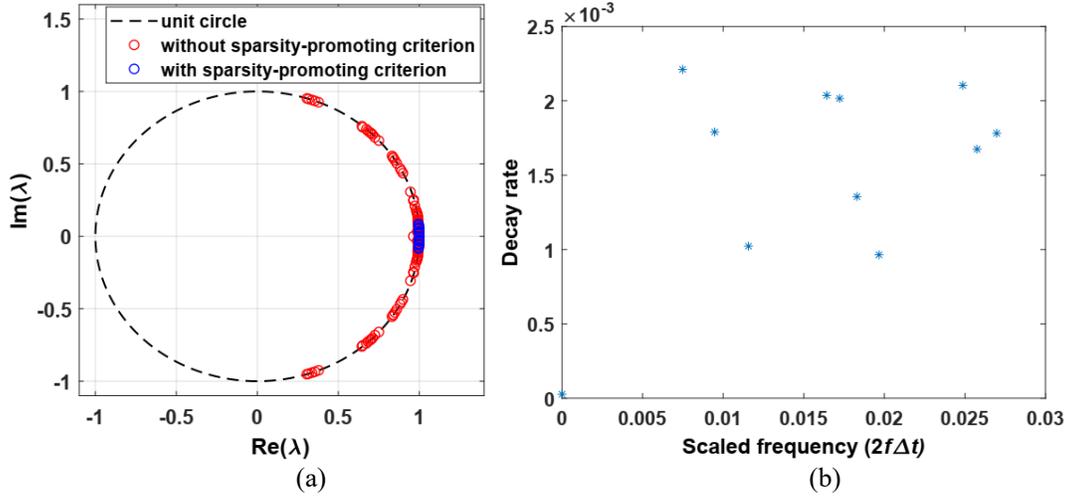

**Fig. 10**: (a) DMD spectrum and (b) decay rate compared to frequency (scaled by $2f\Delta t$) of Test point I

Ultimately, the aeroelastic modes determined from test point I, their frequency (scaled) and damping ratio (ratio of decay rate to the angular frequency) alongside their counterpart from previous literature including HODMD-based method and Least Square fitting of Complex Frequency domain functions (LSCF) method [27–29], are listed in Table 1. The modes identified based on the sparsity-promoting criterion are presented in Table 1, ordered from the lowest to the highest frequencies. Most of the frequencies detected by previous studies [27–29], have been detected by the present framework automatically utilizing sparsity-promoting criterion. Although the damping ratio matches quite well with previous HODMD-based methods, it's somewhat different in comparison to the LSCF-based method. The primary reason behind this is due to the significant amount of noise in the data, and the calculation of the damping ratio is extremely sensitive to noise level. Therefore, it is possible that all the methods (including the present study, previous HODMD-based method, and LSCF) are calculating the damping ratio with a certain amount of error. Moreover, in flutter flight tests, especially with control surface pulses, there is a lot of uncertainty in the damping values of the modes. On the other hand, the only frequencies that were not detected during the present study are marked in red-colored font and are inconsistently detected by the previous studies.

**Table 1** Test point I frequency and damping comparison between the present approach and previous literature

| Frequency (present study) | Frequency (HODMD) (Mendez et al. [28]) | Frequency (LSCF) (Mendez et al. [28]) | Frequency (Clainche et al. [27]) | Frequency (Mendez et al. [29]) | Damping ratio (present study) | Damping ratio (HODMD) (Mendez et al. [28]) | Damping ratio (LSCF) (Mendez et al. [28]) |
|---|---|---|---|---|---|---|---|
| 0.0075 | - | - | - | - | 0.09 | - | - |
| 0.0095 | 0.009 | - | - | - | 0.06 | 0.08 | - |
| 0.0116 | 0.012 | 0.012 | 0.0116 | 0.0116 | 0.03 | 0.03 | 0.05 |
| 0.0164 | 0.016 | - | - | - | 0.04 | 0.05 | - |
| 0.0172 | 0.017 | 0.017 | 0.017 | 0.0173 | 0.04 | 0.04 | 0.09 |
| 0.0183 | 0.018 | - | - | - | 0.03 | 0.05 | - |
| 0.0197 | - | - | - | - | 0.02 | - | - |
| 0.0249 | 0.025 | 0.025 | - | 0.0249 | 0.03 | 0.03 | 0.05 |
| 0.0258 | - | - | 0.0253 | - | 0.02 | - | - |
| 0.027 | 0.027 | 0.027 | 0.0267 | 0.0268 | 0.02 | 0.03 | 0.03 |
| | 0.148 | | | | | | |
| | | | | 0.0789 | | | |

*5.2 Flight Test Point II and III*

The developed framework described above is implemented to analyze the data in test points II and III and is discussed briefly. At first, as discussed in the previous section, an SVD is conducted and rank truncated. After that, the Hankel matrix is created by considering the time-delay criterion and DMD is computed on that Hankel matrix. After convergence is achieved, the sparsity-promoting criterion is implemented to extract the dominant aeroelastic modes. The optimally sparse modes are selected for test points II and III based on the sparsity value and performance loss to obtain the sparsely optimal model as shown in Fig. 11a and 11b respectively.

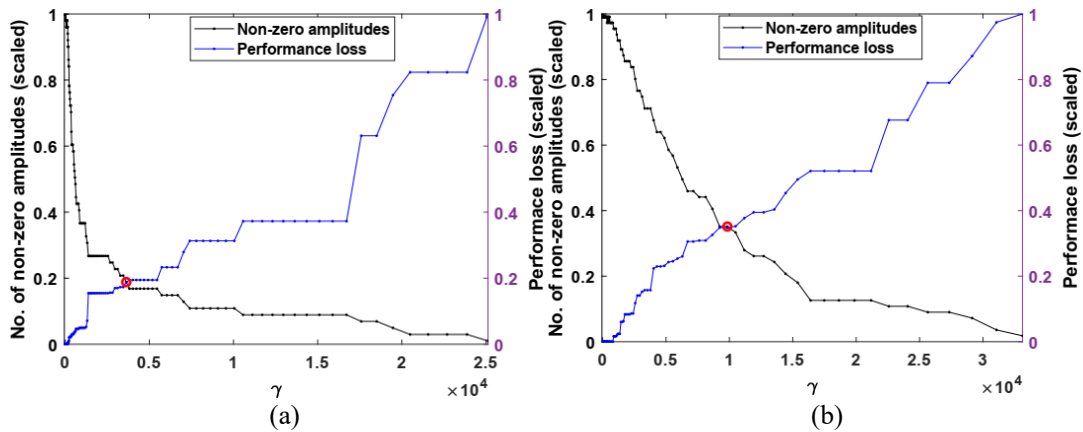

Fig. 11: Selection of optimal sparsity value based on sparsity structure and performance loss for (a) Test point II (b) Test point III

Finally, all the identified modes, their frequencies (scaled), and their damping ratios are listed in Table 2. The modes are ordered from low to high frequencies. 9 modes are identified for test point II, and 17 modes are identified from test point III including extracted damping values. An interesting fact is that the current methodology was able to extract some additional modes with relatively higher frequencies and lower damping values in test point III compared to test points I and II. Although no information regarding the test points (including aircraft speed, altitude, sampling rate, etc.) is known due to the proprietary nature of the dataset, it can be assumed that due to the presence of those additional excited modes with higher frequencies and lower damping, test point III is considerably more unstable compared to test point I and II.

It is to be noted that the computational time for analyzing each test point is around 45 seconds or less on a laptop-class computer, making it an extremely fast and efficient method for accurately detecting the dominant aeroelastic modes from flutter flight test data. Therefore, this reflects the effectiveness of the developed framework and showcases its potential for possible implementation during flutter testing.

Table 2  Test point II and III results

| Test point II | | Test point III | |
|---|---|---|---|
| Frequency | Damping ratio | Frequency | Damping ratio |
| 0.0075 | 0.09 | 0.0096 | 0.02 |
| 0.0093 | 0.06 | 0.011 | 0.024 |
| 0.0116 | 0.03 | 0.0118 | 0.01 |
| 0.0166 | 0.05 | 0.0123 | 0.03 |
| 0.0171 | 0.04 | 0.0137 | 0.01 |
| 0.0182 | 0.02 | 0.0158 | 0.01 |
| 0.0195 | 0.02 | 0.0165 | 0.05 |
| 0.0249 | 0.025 | 0.0173 | 0.01 |
| 0.0268 | 0.02 | 0.0185 | 0.03 |
| - | - | 0.0196 | 0.03 |
| - | - | 0.0223 | 0.01 |
| - | - | 0.0252 | 0.03 |
| - | - | 0.0437 | 0.001 |
| - | - | 0.0453 | 0.001 |
| - | - | 0.0465 | 0.001 |
| - | - | 0.0476 | 0.001 |
| - | - | 0.0518 | 0.002 |

*5.3 Rapid Detection of Aeroelastic Modes Based on Limited Sensor Measurements*

In the case of limited sensor measurements, the developed framework is adjusted by augmenting with a compressed sensing algorithm to extract the dominant aeroelastic modes even if measurements are severely limited (Eq. (25) -(31)). The idea behind compressed sensing is that the full-state measurement $X$ is sparse in a transformation basis $\Psi$, such that $X = \Psi S$, and $Y = CX = C\Psi S$. The measurement matrix $C$ must be incoherent with respect to $\Psi$ so that it preserves the restricted isometric property [54] for the K-sparse vector (K non-zero element in vector $S$), so that,

$$(1 - \delta_k)\|S\|_2^2 \leq \|C\Psi S\|_2^2 \leq (1 + \delta_k)\|S\|_2^2 \qquad (31)$$

Therefore, the sensor selection must be random in order to hold the restrictive isometric property. The choice of measurement matrix $C$ is very important so that it remains incoherent with the transformation basis $\Psi$. In this work, three different types of measurement matrices are considered, namely, uniform random measurement, Gaussian random measurement, and single pixel measurement. Among these, Gaussian random projection is widely applied in the compressed sensing community. If the Gaussian measurement matrix has a dimension of $n \times m$, then each element of the matrix obeys a Gaussian distribution with zero mean and variance $1/n$ independently. The Gaussian measurement matrix is totally random and is uncorrelated with most sparse bases. Therefore, it is generally treated as a universal measurement matrix. In most engineering applications, it is quite common to represent a signal in a generic Fourier or wavelet basis, and single pixel measurements have one key advantage, they are incoherent concerning these generic bases and excite a broadband frequency response [34]. The three different measurement matrices are illustrated in Fig. 12. In this section, the application of these measurement matrices to achieve full-state reconstruction and thereby aeroelastic mode identification is discussed.

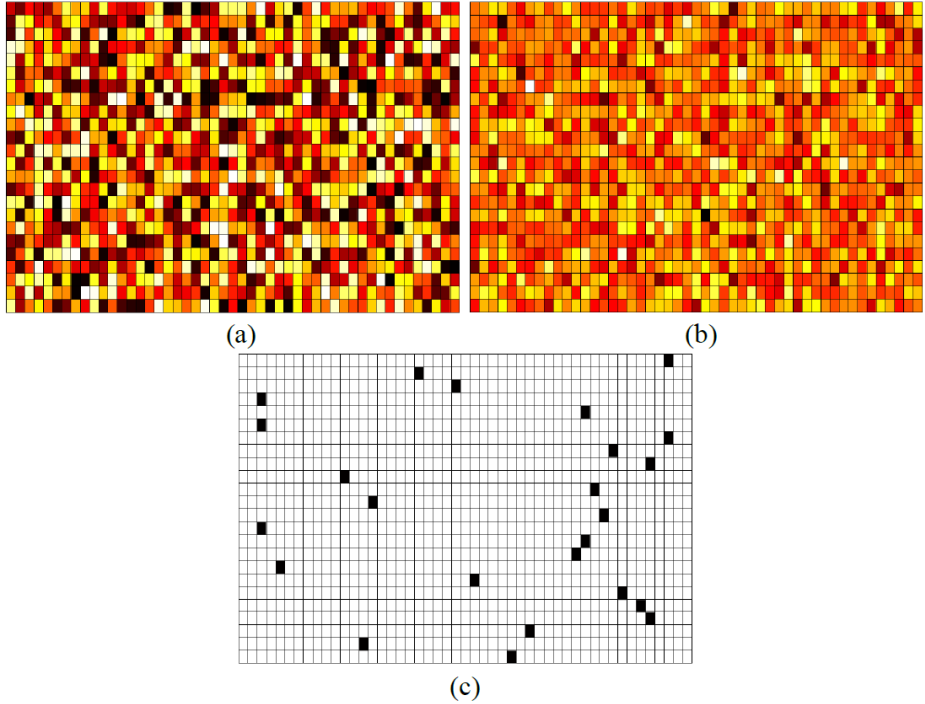

**Fig. 12**: (a) Uniform random projection, (b) Gaussian random projection, (c) Single pixel projection

This section discusses the effectiveness of these three compressibility matrices applied to compressed sensing DMD-based framework in correctly identifying the aeroelastic modes with limited sensor measurements and comparing them with the results obtained from all the sensors (87 in total) that are available. Moreover, the robustness and limitation of the developed algorithm is tested by reducing the sensor number from moderately low to extremely low. For the sake of brevity and to avoid redundancy, only the results from flight test I are analyzed using the compressed sensing-based algorithm developed for extracting the aeroelastic modes from subsampled measurements.

*5.3.1 Uniform random projection*

The framework illustrated in Fig. 6b shows the procedure for extracting the aeroelastic modes based on limited measurements. In this case, a uniform random projection matrix is utilized to randomly select a few sensors from the available 87. To check the efficacy of the algorithm, various number of sensors (20, 15, 10, and 5) was chosen to test the efficacy of the developed framework. Table 3 shows the comparison between the framework using all the sensors available (87) and the compressed sensing DMD-based framework. To preserve brevity, the results obtained by selecting 20 sensors and 5 sensors are listed in Table 3. The frequencies and damping ratios obtained by limited sensor measurements are listed and compared with the frequencies and damping ratios obtained by all (87) sensors with their respective error values. For all the sensor measurements ranging from 20 to 5, 6 out of 10 modes are detected compared to all the sensors. It is noteworthy that the natural frequencies of the 6 modes are detected with satisfactory accuracy within an error margin of 0 to 5%. However, out of these six modes, 3 of them have the damping ratio exactly right but the other 3 capture the damping ratio with an error ranging from 16% to 50% as listed in Table 3 for all the measurements. It is to be noted that, computations based on 20 and 15 sensor measurements extracted three extra natural frequencies, and the rest detected two extra natural frequencies (possibly a numerical artifact) which were not present while using all the sensors for computations and are marked in a red-colored font. It is quite interesting that while decreasing the measurement from 20 to 5 sensors, the accuracy does not get compromised by a significant amount.

Table 3  Comparison of aeroelastic modes for all Vs limited sensors (Uniform random projection)

| All (87) sensors Frequency | All (87) sensors Damping ratio | Limited (20) sensors Frequency | Limited (20) sensors Frequency Error (%) | Limited (20) sensors Damping Ratio | Limited (20) Damping Ratio Error (%) | Limited (5) sensors Frequency | Limited (5) sensors Frequency Error (%) | Limited (5) sensors Damping Ratio | Limited (5) Damping Ratio Error (%) |
|---|---|---|---|---|---|---|---|---|---|
| 0.0075 | 0.09 | - | - | - | - | - | - | - | - |
| 0.0095 | 0.06 | 0.0099 | 4.21 | 0.07 | 16.67 | 0.01 | 5.26 | 0.07 | 16.67 |
| 0.0116 | 0.03 | 0.0116 | 0 | 0.02 | 33.33 | 0.0115 | 0.86 | 0.02 | 33.33 |
| - | - | 0.0153 | - | 0.16 | - | 0.0153 | - | 0.18 | - |
| 0.0164 | 0.04 | 0.0168 | 2.44 | 0.05 | 25 | 0.017 | 3.66 | 0.05 | 25 |
| 0.0172 | 0.04 | 0.0175 | 1.74 | 0.04 | 0 | 0.0176 | 3.53 | 0.04 | 0 |
| 0.0183 | 0.03 | - | - | - | - | - | - | - | - |
| 0.0197 | 0.02 | - | - | - | - | - | - | - | - |
| - | - | 0.0227 | - | 0.01 | - | 0.0227 | - | 0.01 | - |
| 0.0249 | 0.03 | 0.0248 | 0.4 | 0.03 | 0 | 0.0248 | 0.4 | 0.03 | 0 |
| 0.0258 | 0.02 | - | - | - | - | - | - | - | - |
| 0.027 | 0.02 | 0.0268 | 0.74 | 0.02 | 0 | 0.0268 | 0.74 | 0.02 | 0 |
| - | - | 0.08 | - | 0.02 | - | - | - | - | - |

*5.3.2 Gaussian random projection*

While the results obtained by incorporating uniform random projection as the compressibility matrix are satisfactory, using Gaussian random projection drastically improves the results. All the sensor measurements from 20 to 5 were able to extract the aeroelastic modes successfully as shown in Table 4. The frequencies are captured within an error margin of 0 to 1% for all measurement values. Moreover, the damping ratios attributed to 20 sensor measurements are exactly equal compared to their counterparts with all the sensors except for just one mode with a 33% error. On the other hand, in the case of 5 sensor measurements, two out of ten modes have damping ratios with an error value of 33%, and the rest are the same compared to all sensor measurements. It is noteworthy that the results obtained even with only 5 sensor measurements are highly satisfactory. Nevertheless, all the sensor measurements show impeccable results.

**Table 4** Comparison of aeroelastic modes for all Vs limited sensors (Gaussian random projection)

| All (87) sensors Frequency | All (87) sensors Damping ratio | Limited (20) sensors Frequency | Limited (20) sensors Frequency Error (%) | Limited (20) sensors Damping ratio | Limited (20) sensors Damping Ratio Error (%) | Limited (5) sensors Frequency | Limited (5) sensors Frequency Error (%) | Limited (5) sensors Damping ratio | Limited (5) sensors Damping Ratio Error (%) |
|---|---|---|---|---|---|---|---|---|---|
| 0.0075 | 0.09 | 0.0074 | 1.33 | 0.09 | 0 | 0.0075 | 0 | 0.09 | 0 |
| 0.0095 | 0.06 | 0.0094 | 1.05 | 0.06 | 0 | 0.0095 | 0 | 0.06 | 0 |
| 0.0116 | 0.03 | 0.0116 | 0 | 0.03 | 0 | 0.0116 | 0 | 0.03 | 0 |
| 0.0164 | 0.04 | 0.0165 | 0.61 | 0.04 | 0 | 0.0164 | 0 | 0.04 | 0 |
| 0.0172 | 0.04 | 0.0173 | 0.58 | 0.04 | 0 | 0.0172 | 0 | 0.04 | 0 |
| 0.0183 | 0.03 | 0.0183 | 0 | 0.02 | 33.33 | 0.0183 | 0 | 0.02 | 33.33 |
| 0.0197 | 0.02 | 0.0197 | 0 | 0.02 | 0 | 0.0197 | 0 | 0.02 | 0 |
| 0.0249 | 0.03 | 0.0249 | 0 | 0.03 | 0 | 0.0248 | 0.4 | 0.03 | 0 |
| 0.0258 | 0.02 | 0.0261 | 1.16 | 0.02 | 0 | 0.0259 | 0.39 | 0.03 | 33.33 |
| 0.027 | 0.02 | 0.0271 | 0.37 | 0.02 | 0 | 0.0271 | 0.37 | 0.02 | 0 |

*5.3.3 Single pixel projection*

The final compressibility matrix considered for this work is the single pixel projection matrix. Due to its practicality and ease of use, sampling at random point locations is appealing when individual measurements are expensive. Single pixel measurements are used prevalently in engineering applications. Moreover, it is advantageous because single pixel measurements are incoherent with most generic bases, such as Fourier and wavelet bases. As shown in Table 5, the results obtained are strikingly good. For all measurement values, the compressed sensing-based framework can detect all the frequencies with high accuracy and within an error margin between 0 to 1.5%. The damping ratios obtained from 20 sensor measurements are excellent, except for just one mode with 33% error just like the previous case with random Gaussian projection. Although further reduction in sensor number holds up well for all the frequencies, three out of 10 damping ratios have an error value of 11% to 50%, the rest are exactly equal compared to all sensor measurements as shown in Table 5.

Table 5  Comparison of aeroelastic modes for all Vs limited sensors (Single pixel projection)

| All (87) sensors Frequency | All (87) sensors Damping ratio | Limited (20) sensors Frequency | Limited (20) sensors Frequency Error (%) | Limited (20) sensors Damping ratio | Limited (20) Damping Ratio Error (%) | Limited (5) sensors Frequency | Limited (5) sensors Frequency Error (%) | Limited (5) sensors Damping ratio | Limited (5) Damping Ratio Error (%) |
|---|---|---|---|---|---|---|---|---|---|
| 0.0075 | 0.09 | 0.0076 | 1.33 | 0.09 | 0 | 0.0075 | 0 | 0.08 | 11.11 |
| 0.0095 | 0.06 | 0.0095 | 0 | 0.06 | 0 | 0.0095 | 0 | 0.06 | 0 |
| 0.0116 | 0.03 | 0.0116 | 0 | 0.03 | 0 | 0.0116 | 0 | 0.03 | 0 |
| 0.0164 | 0.04 | 0.0165 | 0.61 | 0.04 | 0 | 0.0165 | 0.61 | 0.04 | 0 |
| 0.0172 | 0.04 | 0.0173 | 0.58 | 0.04 | 0 | 0.0171 | 0.58 | 0.04 | 0 |
| 0.0183 | 0.03 | 0.0183 | 0 | 0.03 | 0 | 0.0182 | 0.55 | 0.02 | 33.33 |
| 0.0197 | 0.02 | 0.0196 | 0.51 | 0.02 | 0 | 0.0194 | 1.52 | 0.01 | 50 |
| 0.0249 | 0.03 | 0.0248 | 0.4 | 0.03 | 0 | 0.0249 | 0 | 0.03 | 0 |
| 0.0258 | 0.02 | 0.0257 | 0.39 | 0.03 | 33.33 | 0.0255 | 1.16 | 0.02 | 0 |
| 0.027 | 0.02 | 0.027 | 0 | 0.02 | 0 | 0.027 | 0 | 0.02 | 0 |

All three abovementioned projection matrices performed satisfactorily in obtaining the dominant aeroelastic modes from flutter flight test data. However, it needs to be mentioned that Gaussian random projection and single pixel measurement outperform the uniform random projection method, while the Gaussian random delivered the most accurate result. Furthermore, the computational time taken for each test point with measurement values ranging from 5 to 20 varies between 32 to 38 seconds. This shows the potential for this method for possible real-time implementation during flutter flight tests.

## 6. Conclusions

This study showcases an innovative approach for rapid detection of aeroelastic modes from flutter flight test data based on limited sensor measurements. First, a time-delay embedded DMD-based framework for rapid detection of the aeroelastic modes has been developed in conjunction with an RPCA algorithm, and sparsity-promoting criterion for optimally sparse mode selection. The developed framework removes part of the noise and most outliers by using the RPCA filtering technique, and the dominant modes are automatically extracted by utilizing a sparsity-promoting criterion for mode selection. The framework is tested with anonymized flutter flight test data and validated against previous literature with the HODMD-based method and LSCF method. Furthermore, the developed framework is modified by augmenting with a compressed sensing DMD algorithm for rapid detection of aeroelastic modes based on subsampled measurements.

The efficacy and robustness of the framework have been tested thoroughly by considering three different measurement matrices, namely, uniform random, gaussian random, and single pixel projection matrix. Although all three performed satisfactorily, Gaussian random and single pixel projection methods outperformed uniform random by detecting all the frequencies with very high accuracy. A significantly low number of sensor measurements are taken, starting from 20 and up to 5, out of a total of 87 sensors to test the efficacy of the developed framework. Although, the performance of the Gaussian random and single pixel projection methods in capturing the damping ratios are the same while considering 20 sensor measurements, dropping down the sensor value to 5 shows that Gaussian random is slightly superior compared to all the three projection methods considered in this study. Furthermore, the computational time for each test point with the different measurement values considered with different projection methods varies from 32 to 38 seconds while computing in a laptop class computer. This shows the potential and the possibility of this developed framework for real-time implementation during flutter flight tests.

## Acknowledgments

The authors would like to thank the Defence Science and Technology Group of the Australian Department of Defence for providing partial funding support for this research.